\documentclass[showpacs,10pt,twocolumn,prb]{revtex4-1}
\usepackage{amsmath}
\usepackage{amssymb}
\usepackage{graphics}
\usepackage{epsfig}

\begin{document}

\title{Effects of excess Fe on upper critical field and magnetotransport in
Fe$_{1+y}$(Te$_{1-x}$S$_{x}$)$_{z}$}
\author{Hechang Lei,$^{1}$ Rongwei Hu,$^{1,\ast }$ E. S. Choi,$^{2}$ J. B.
Warren,$^{3}$ and C. Petrovic$^{1}$}
\affiliation{$^{1}$Condensed Matter Physics and Materials Science Department, Brookhaven
National Laboratory, Upton, New York 11973, USA}
\affiliation{$^{2}$NHMFL/Physics, Florida State University, Tallahassee, Florida 32310,
USA}
\affiliation{$^{3}$Instrumentation Division, Brookhaven National Laboratory, Upton, New
York 11973, USA}
\date{\today}

\begin{abstract}
We have investigated the upper critical field anisotropy and
magnetotransport properties of Fe$_{1.14(1)}$Te$_{0.91(2)}$S$_{0.09(2)}$
single crystals in stable magnetic fields up to 35 T. The results show that $%
\mu _{0}H_{c2}(T)$ along the c axis and in the ab-plane exhibit saturation
at low temperatures. The anisotropy of $\mu _{0}H_{c2}(T)$ decreases with
decreasing temperature, becoming nearly isotropic for T$\rightarrow $0. Our
analysis indicates that the spin-paramagnetic pair-breaking with
spin-orbital scattering is responsible for the behavior of $\mu
_{0}H_{c2}(T) $. Furthermore, from analysis of the normal state properties,
we show evidence that the excess Fe acting as Kondo-type impurities is a key
factor determining the normal and superconducting state physical properties.
\end{abstract}

\maketitle

\section{Introduction}

Iron-based superconductors have generated a great deal of interests due to
exotic physical and chemical properties such as high transition temperature $%
T_{c}$ (above 50 K) in layered structure without copper oxygen planes, spin
fluctuation spectrum dominated by two dimensional incommensurate excitations
comparable to high-$T_{C}$ cuprates and multiorbital physics with active
spin, charge and orbital degrees of freedom.\cite{Kamihara}$^{-}$\cite{Lee
CC} Simple binary FeSe$_{x}$, Fe(Te$_{1-x}$Se$_{x}$)$_{z}$, and Fe$_{1+y}$(Te$%
_{1-x}$S$_{x}$)$_{z}$\cite{Hsu FC}$^{-}$\cite{Mizuguchi} share common characteristics with other
iron-based superconductors: a square-planar lattice of Fe with tetrahedral
coordination and similar Fermi surface topology.\cite{Subedi} On the other
hand they exhibit some distinctive features such as the absence of charge
reservoir, significant pressure effect\cite{Mizuguchi2} and strongly
magnetic excess Fe in Fe(2) site providing local moments that are expected
to persist even if the antiferromagnetic order is suppressed by doping or
pressure.\cite{Zhang LJ} Furthermore, superconductivity in Fe$_{1+y}$(Te$%
_{1-x}$S$_{x}$)$_{z}$ develops from nonmetallic conductivity which is different
from metallic resistivity above T$_{C}$ in all other iron based
superconductors.\cite{Mizuguchi}$^{,}$\cite{Hu RW}

There are two remarkable common characteristics in $\mu _{0}H_{c2}$-$T$
phase diagram of iron-based superconductors. In ternary and quaternary iron
pnictide superconductors (122 and 1111 systems) $\mu _{0}H_{c2,c}(T)$ shows
pronounced upturn or positive temperature curvature far below $T_{c}$
without saturation. In contrast, $\mu _{0}H_{c2,ab}(T)$ exhibits a downturn
curvature with decreasing temperature.\cite{Yuan HQ}$^{,}$\cite{Jaroszynski}%
\ The former can be explained by two band theory with high (1111) or low
(122 systems) intraband diffusivity ratio of electron band to hole band and
the latter is commonly ascribed to the spin-paramagnetic effect.\cite%
{Jaroszynski}$^{-}$\cite{Kano}

Here we report comprehensive study of the upper critical field anisotropy
and magnetotransport properties of Fe$_{1.14(1)}$(Te$_{0.91(2)}$S$_{0.09(2)}$%
)$_{z}$ single crystals in stable magnetic fields up to 35 T. We observe
that enhanced spin-paramagnetic effect is dominant in both $\mu
_{0}H_{c2,c}(T)$ and $\mu _{0}H_{c2,ab}(T)$. We conclude that the root cause
of that enhancement and the anomalous normal state electronic transport
properties is the existence of excess Fe(2) iron. As opposed to 122 and 1111
iron pnictide superconductors derived from stoichiometric Ba(Sr)Fe$_{2}$As$%
_{2}$ and LaOFeAs parent compounds, the width of material formation and
subtle iron stoichiometry is rather important in superconductors derived
from Fe$_{1+y}$Te.

\section{Experiment}

Single crystals of Fe(Te,S) were grown by self flux method and their crystal
structure was analyzed in the previous report.\cite{Hu RW} The elemental and
microstructure analysis on particular crystal used in this study showed Fe$%
_{1.14(1)}$(Te$_{0.91(2)}$S$_{0.09(2)}$)$_{z}$ stoichiometry and will be
denoted as S-09 in the following for brevity. Electrical transport
measurements were performed using a four-probe configuration with current
flowing in the ab-plane of tetragonal structure in dc magnetic fields up to
9 T in a Quantum Design PPMS-9 from 1.8 to 200 K and up to 35 T in an Oxford
Heliox cryostat with resistive magnet down to 0.3 K at the National High
Magnetic Field Laboratory (NHMFL) in Tallahassee, FL.

\section{Results and Discussions}

Fig.\ 1(a,b) shows the temperature dependent ab-plane electrical resistivity
$\rho _{ab}(T)$ of S-09 below 15 K in magnetic fields from 0 to 9 T for H$%
\Vert $ab and H$\Vert $c. With increasing magnetic fields, the resistivity
transition widths are slightly broader. The onset of superconductivity
shifts to lower temperatures gradually for both magnetic field directions,
but the trend is more obvious for H$\Vert $c than H$\Vert $ab. The shape and
broadening of $\rho _{ab}(T)$ for H$\Vert $c is comparable to 122-system,%
\cite{Wang ZS} but quite different from 1111-system\cite{Lee HS} where it
was explained by the vortex-liquid state similar to cuprates.\cite{Pribulova}%
$^{-}$\cite{Safar} Hence, it can be concluded that the vortex-liquid state
region is narrower or even absent in S-09. This is similar to Fe$_{1+y}$(Te$%
_{1-x}$Se$_{x}$)$_{z}$.\cite{Lei HC}

\begin{figure}[tbp]
\centerline{\includegraphics[scale=0.8]{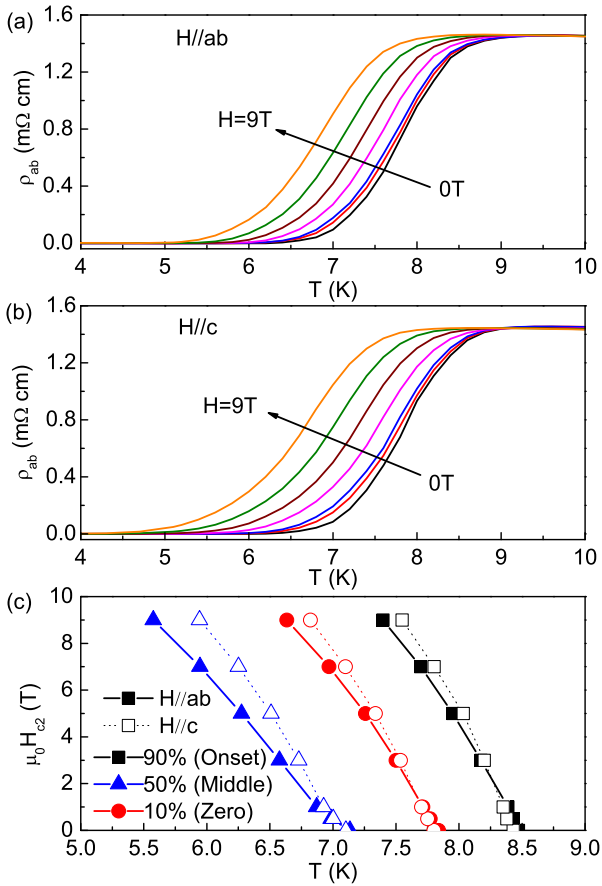}}
\vspace*{-0.3cm}
\caption{(a) and (b) Temperature
dependence of $\protect\rho _{ab}$($T$) of S-09 at fixed fields (0, 0.5, 1,
3, 5, 7, 9 T) for H$\Vert $ab plane and H$\Vert $c axis below 15 K,
respectively. (c) Temperature dependence of the resistive upper critical
field $\protect\mu _{0}$H$_{c2}$($T$) corresponding three defined
temperatures at low fields.}
\end{figure}

\begin{figure}[tbp]
\centerline{\includegraphics[scale=0.8]{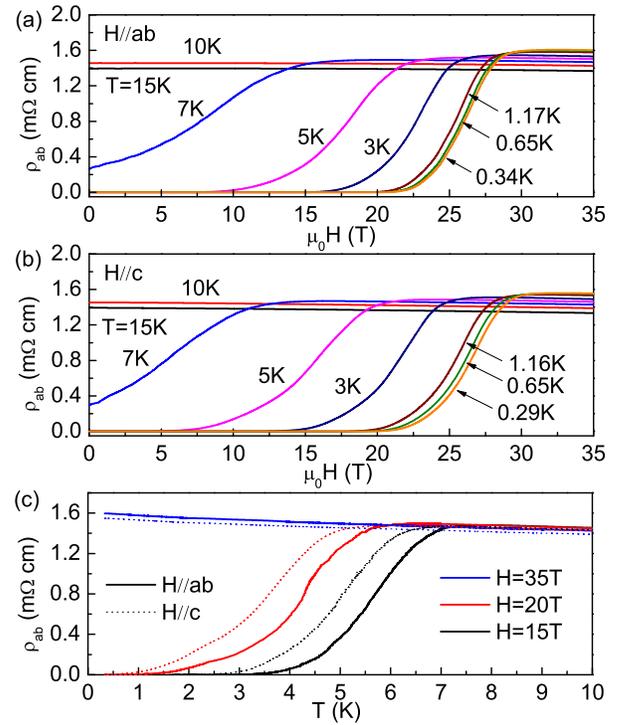}}
\vspace*{-0.3cm}
\caption{Field dependence of $\protect\rho _{ab}(H)$ measured at various temperatures in dc magnetic fields up to
35 T for (a) H$\Vert $ab and (b) H$\Vert $c. (c) Temperature dependence of $\protect\rho _{ab}(T)$ at high magnetic fields from 15 to 35 T (15, 20 and
35 T).}
\end{figure}

The upper critical field $\mu _{0}H_{c2}(T)$ corresponding to temperatures
where the resistivity drops to 90\%, 50\% and 10\% of the normal state
resistivity $\rho _{n,ab}(T,H)(T_{c,onset})$ is shown in Fig. 1(c). The
normal-state resistivity $\rho _{n,ab}(H,T)$ was determined by linearly
extrapolating the normal-state behavior above the onset of superconductivity
in $\rho _{ab}(T)$ curves (same as for $\rho _{ab}(H)$ curves). The slope of
$\mu _{0}H_{c2}(T_{c})$ obtained from linear fitting the curves of $\mu
_{0}H_{c2}(T)$ near $T_{c}$ for all defined temperatures are listed in Table
1. The values of orbital pair breaking field $\mu _{0}H_{c2}^{\ast }(0)$
corresponding to the conventional one-band Werthamer-Helfand-Hohenberg (WHH)
theory\cite{Werthamer} $\mu _{0}H_{c2}^{\ast }(0)$=-0.693$(d\mu
_{0}H_{c2}/dT)_{Tc}Tc$ are also listed in Table 1.

\begin{table*}[tbp] \centering%
\caption{($d\mu _{0}H_{c2}/dT)_{Tc}$ and derived $\mu _{0}H_{c2}^{*}(0)$
data at three defined temperatures using WHH formula. $\mu _{0}H_{c2,ab}^{*}(0)$ and $\mu _{0}H_{c2,c}^{*}(0)$
are the ab-plane  and c-axis orbital-limited upper critical fields at T=0 K.}%
\begin{tabular}{cccccc}
\hline\hline
Fe$_{1.14(1)}$Te$_{0.91(2)}$S$_{0.09(2)}$ & $T_{C}$ & $(d\mu
_{0}H_{c2}/dT)_{Tc}$, H$\parallel $ab & $(d\mu _{0}H_{c2}/dT)_{Tc}$, H$%
\parallel $c & $\mu _{0}H_{c2,ab}^{\ast }(0)$ & $\mu _{0}H_{c2,c}^{\ast }(0)$
\\
& (K) & (T/K) & (T/K) & (T) & (T) \\ \hline
Onset & 8.47 & 12.82 & 8.44 & 75.25 & 49.54 \\
Middle & 7.84 & 10.18 & 8.21 & 55.31 & 44.61 \\
Zero & 7.14 & 8.58 & 6.10 & 42.45 & 30.18 \\ \hline\hline
\end{tabular}%
\label{TableKey}%
\end{table*}%

Superconductivity is suppressed by increasing magnetic field up to 35 T and
the transition of $\rho _{ab}(H)$ curves are shifted to lower magnetic
fields at higher measuring temperature (Fig. 2(a,b)). At 0.3 K, the lowest
temperature of our measurement we observe no superconductivity up to 35 T
for both crystallographic directions, indicating that the upper critical
field $\mu _{0}H_{c2}(0)$ of Fe$_{1+y}$(Te$_{1-x}$S$_{x}$)$_{z}$ with same
doping level is lower than that of Fe$_{1+y}$(Te$_{1-x}$Se$_{x}$)$_{z}$.\cite%
{Lei HC} Fig. 2(c) shows the temperature dependence of resistivity at high
magnetic fields. The superconductivity above 0.3 K is suppressed when $\mu
_{0}H$=35 T, irrespective of the direction of field, consistent with the
results of $\rho _{ab}(H)$ measurement. The superconducting transition
widths are only slightly broader even at 20 T, indicating that the
vortex-liquid state in S-09 is narrow not only in low field high temperature
but also in high field low temperature region.

\begin{figure}[tbp]
\centerline{\includegraphics[scale=0.8]{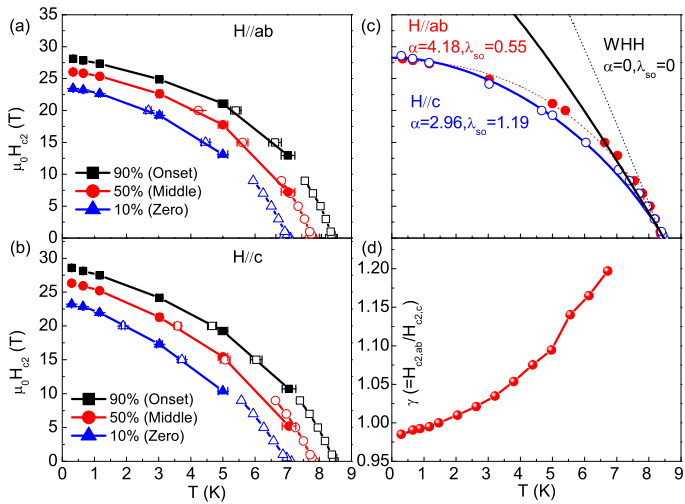}}
\vspace*{-0.3cm}
\caption{Temperature dependence of the
resistive upper critical field $\protect\mu _{0}H_{c2}(T)$ of S-09 for (a) H$\Vert$ab and (b) H$\Vert$c derived from $\protect\rho _{ab}$($T$) (open
symbols) and $\protect\rho _{ab}$($H$) (closed symbols) curves. (c) Analysis
of $\protect\mu _{0}H_{c2,onset}(T)$ for H$\Vert$ab (closed red circles)
and H$\Vert$c (open blue circles) using the WHH theory with and without
spin-paramagnetic effect and spin-orbital scattering (dotted black and red
curve for H$\Vert $ab and solid black and blue curve for H$\Vert $c,
respectively). (d) The anisotropy of the upper critical field, $\protect\gamma $=$H_{c2,ab}(T)/H_{c2,c}(T)$, as a function of temperature.}
\end{figure}

From the results of $\rho _{ab}(H)$ and $\rho _{ab}(T)$ at low and high
field, we construct the $\mu _{0}H_{c2}(T)$-$T$ phase diagram (Fig. 3(a,b)).
There is a linear increase in $\mu _{0}H_{c2}(T)$ with decreasing
temperature near $T_{c}$ and a saturation trend away from $T_{c}$
irrespective of field direction, similar to Fe$_{1+y}$(Te$_{1-x}$Se$_{x}$)$%
_{z}$.\cite{Lei HC} This is different from 1111- and 122-system, which
exhibit upturn or linear behavior at low temperature for $\mu _{0}H_{c2,c}(T)
$ ascribed to two band effect.\cite{Jaroszynski}$^{,}$\cite{Yuan HQ} The $%
\mu _{0}H_{c2,onset}(0)$ is about 28T for both field directions. This is
much smaller than the values predicted by WHH formalism by only considering
orbital pair-breaking effect (Table 1, Fig. 3(c) black lines).

In what follows we consider the contribution of spin-paramagnetic effect and
its origin. Only the $\mu _{0}H_{c2,onset}(T)$ were chosen for further
analysis.\cite{Jaroszynski}$^{,}$\cite{Fuchs} The effects of Pauli spin
paramagnetism and spin-orbit scattering were included in the WHH\ theory
through the Maki parameters $\alpha $ and $\lambda _{so}$.\cite{Werthamer}$%
^{,}$\cite{Maki} We found it necessary to introduce $\lambda _{so}\neq 0$ in
$\mu _{0}H_{c2}(T)$ fits, unlike for Fe$_{1+y}$(Te$_{1-x}$Se$_{x}$)$_{z}$.%
\cite{Lei HC}$^{,}$\cite{Fuchs} The results (Fig. 3(c)) indicate that the
spin-paramagnetic effect is the dominant pair-breaking mechanism in S-09 for
both H$\Vert $ab and H$\Vert $c. The calculated zero-temperature
Pauli-limited field\cite{Maki} $H_{p}(0)$=$\sqrt{2}H_{c2}^{\ast }(0)$/$%
\alpha $ using $\alpha $ obtained from $\mu _{0}H_{c2}(T)$ fits and
zero-temperature coherence length $\xi (0)$ estimated with Ginzburg-Landau
formula $\mu _{0}H_{c2}(0)$=$\Phi _{0}$/2$\pi \xi (0)$ (where $\Phi _{0}$%
=2.07$\times $10$^{-15}$ Wb) are listed in Table 2. The $\mu _{0}H_{p}(0)$
is smaller than $\mu _{0}H_{c2}(0)$ due to the large value of $\alpha $.
Since Fe$_{1+y}$Te$_{1-x}$S$_{x}$ superconductors are in the dirty limit,%
\cite{Kim H} the Fulde-Ferrell-Larkin-Ovchinnikov (FFLO) state at high
fields is unlikely because short mean free path will remove any momentum
anisotropy.\cite{Fulde}$^{,}$\cite{Larkin} The two-band theory,\cite%
{Gurevich} , which is applicable to the 1111-system, did not yield
satisfactory fits (not shown here).

\begin{figure}[tbp]
\centerline{\includegraphics[scale=0.8]{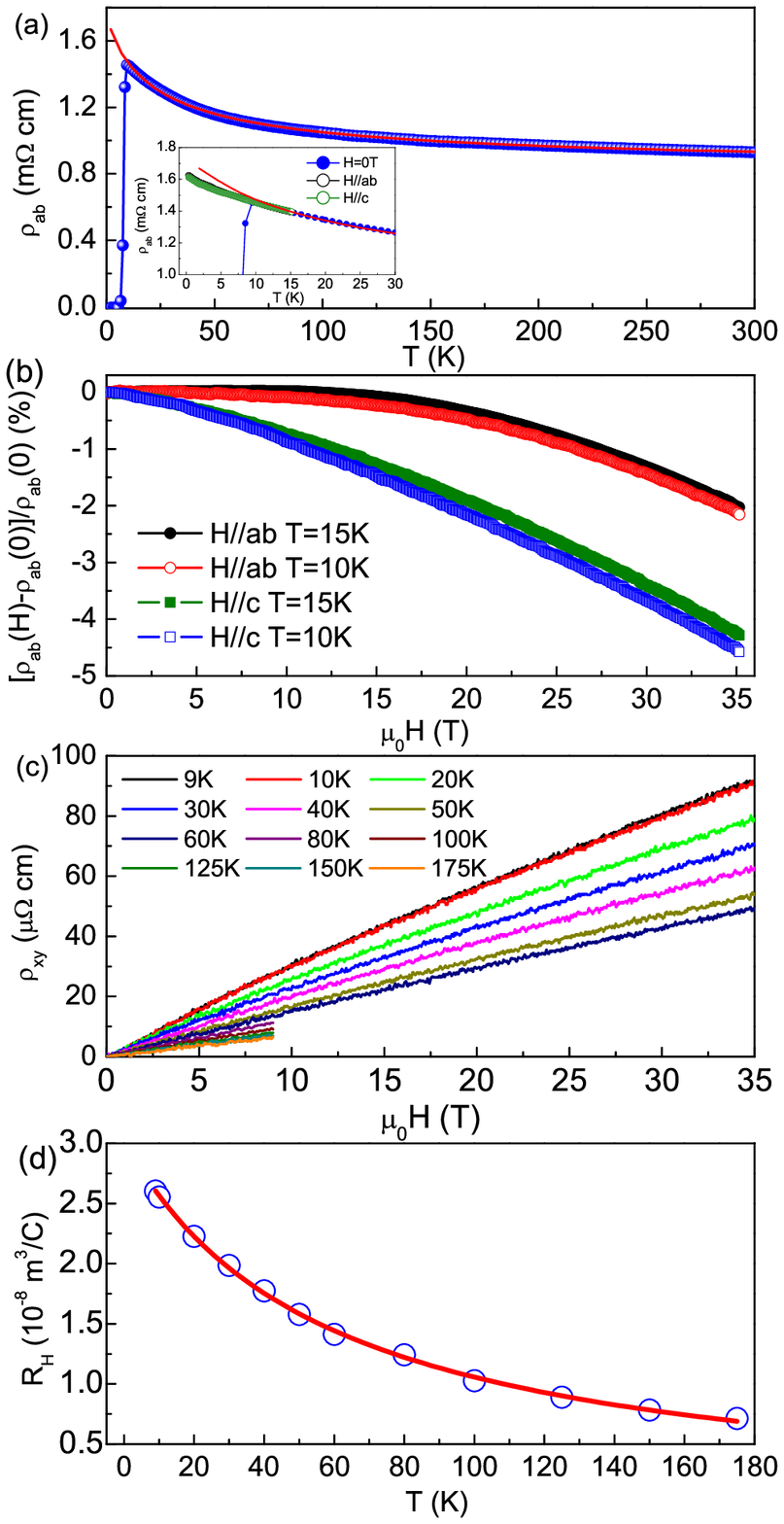}}
\vspace*{-0.3cm}
\caption{(a) Temperature dependence of $\protect\rho _{ab}$($T$) at zero field from 1.8-300 K and fitted curve using
Hamann's equation. Inset of (a) shows the $\protect\rho _{ab}$($T$) at 0 and
35 T for H$\Vert $ab and H$\Vert $c (deducting the MR effects) at low
temperature. (b) Field dependence of MR at different temperatures up to 35 T
at 10 and 15 K. (c) Hall resistivity $\protect\rho _{xy}$ vs. magnetic field
at various temperatures, and the data above 60 K are obtained in PPMS system
up to 9 T. (d) Temperature dependence of Hall coefficient R$_{H}$ determined
from linear fitting $\protect\rho _{xy}(H)$ data and fitting curve using the
formula described in the text.}
\end{figure}

\begin{table*}[tbp] \centering%
\caption{Superconducting parameters of S-09 obtained from the analysis of $\mu_{0}H_{c2,onset}(T)$.
 $\mu_{0}H_{c2}^{*}(0)$,  $\mu_{0}H_{p}(0)$ and $\mu_{0}H_{c2}(0)$ are the zero-temperature orbital-, Pauli-limited and fitted upper critical fields, respectively. $\alpha$ and $\lambda_{so}$ are the fitted Maki parameter and spin-orbital scattering constant, respectively. $\xi_{ab}(0)$ and $\xi_{c}(0)$ are the c-axis and ab-plane zero-temperature coherence length calculated using $\mu_{0}H_{c2}(0)$, respectively.}%
\begin{tabular}{cccccccccccc}
\hline\hline
$\mu _{0}H_{c2,ab}^{\ast }(0)$ & $\mu _{0}H_{c2,c}^{\ast }(0)$ & $\mu
_{0}H_{p,ab}(0)$ & $\mu _{0}H_{p,c}(0)$ & $\mu _{0}H_{c2,ab}(0)$ & $\mu
_{0}H_{c2,c}(0)$ & $\alpha _{H\Vert ab}$ & $\alpha _{H\Vert c}$ & $\lambda
_{so,H\Vert ab}$ & $\lambda _{so,H\Vert c}$ & $\xi _{ab}(0)$ & $\xi _{c}(0)$
\\
(T) & (T) & (T) & (T) & (T) & (T) &  &  &  &  & (nm) & (nm) \\ \hline
75.25 & 49.54 & 25.46 & 23.67 & 27.83 & 28.28 & 4.18 & 2.96 & 0.55 & 1.19 &
3.41 & 3.42 \\ \hline\hline
\end{tabular}%
\label{TableKey copy(1)}%
\end{table*}%

Temperature dependence of anisotropy of $\mu _{0}H_{c2}(T)$, $\gamma ($=$%
H_{c2,ab}(T)/H_{c2,c}(T))$, is shown in Fig. 3(d) as a function of
temperature $T$. The value of $\gamma $ for S-09 is smaller than that of Fe$%
_{1+y}$(Te$_{1-x}$Se$_{x}$)$_{z}$ at high temperature\cite{Lei HC}$^{,}$\cite%
{Fang MH} and it decreases gradually to 1 with decreasing temperature.
Values of $\gamma $ decrease to less than 1 below $T$=1 K, which has also
been observed in Fe$_{1+y}$(Te$_{0.6}$Se$_{0.4}$)$_{z}$.\cite{Lei HC}$^{,}$%
\cite{Fang MH}

Why are there large Maki parameter $\alpha $ and non-zero $\lambda _{so}$?
First, the Maki parameter can be enhanced due to disorder.\cite{Fuchs}$^{,}$%
\cite{Fuchs2} In this system, disorder can be induced by Te(S)
substitution/vacancies and excess Fe in Fe(2) site, resulting in the
enhancement of spin-paramagnetic effect. Second, according to the expression
of $\mu _{0}H_{p}(0)$ with strong coupling correction considering
electron-boson and electron-electron interaction:\cite{Fuchs}$^{,}$\cite%
{Orlando}$^{,}$\cite{Schossmann} $\mu _{0}H_{p}(0)$=$1.86(1+\lambda
)^{\varepsilon }\eta _{\Delta }\eta _{ib}(1-I)$, where $\eta _{\Delta }$
describes the strong coupling intraband correction for the gap, $I$ is the
Stoner factor $I$=$N(E_{F})J$, $N(E_{F})$ is the electronic density of
states (DOS) per spin at the Fermi energy level $E_{F}$, $J$ is an effective
exchange integral, $\eta _{ib}$ is introduced to describe phenomenologically
the effect of the gap anisotropy, $\lambda $ is electron-boson coupling
constant and $\varepsilon $=0.5 or 1. Since the $N(E_{F})$ of FeS is larger
than that of FeSe,\cite{Subedi} it is likely that the $N(E_{F})$ of Fe$_{1+y}
$(Te$_{1-x}$S$_{x}$)$_{z}$ is larger than that of Fe$_{1+y}$(Te$_{1-x}$Se$%
_{x}$)$_{z}$ with same doping content, which will lead to the larger $\alpha
$. It is consistent with the previous reported result.\cite{Lei HC} Third, $%
\mu _{0}H_{p}(0)$ can be decreased, i.e. larger $\alpha $, if the Stoner
factor increases via enhancement of $J$ by Ruderman-Kittel-Kasuya-Yosida
(RKKY) interaction between local magnetic moments of Fe(2) with itinerant
electrons. We expect low content of S doping to have small effect on high $%
N(E_{F})$.\cite{Subedi}$^{,}$\cite{Zhang LJ} On the other hand, large $%
\lambda _{so}$ can also be explained via increasing Kondo-type scattering
from excess Fe, consistent with the definition of $\lambda _{so}$, which is
proportional to the spin-flip scattering rate.\cite{Werthamer}$^{,}$\cite%
{Maki}

In order to confirm that the excess Fe can be seen as the Kondo-type
impurity, in the next section we study the normal state properties
systematically. Fig.\ 4(a) shows the temperature dependence of $\rho _{ab}(T)
$ in zero field from 1.8 K to 300 K. As seen from the data, S-09 exhibits a
non-metallic resistivity behavior in normal state, in agreement with
measurements on polycrystals.\cite{Mizuguchi} Similar behavior has also been
observed in FeTe with low Se content doping,\cite{Yeh KW}$^{,}$\cite{Sales}$%
^{,}$\cite{Liu TJ} ascribed to two dimensional (2D) weak localization.\cite%
{Liu TJ} However, our analysis indicates clearly that $\rho _{ab}(T)$ can
originate from Kondo-type scattering due to excess Fe. It can be seen
clearly that the normal state resistivity at zero field satisfies Hamann's
equation perfectly (Fig. 4(a)): $\rho $=$\rho _{imp}$+$\rho _{0}$[1-ln($T$/$%
T_{K}$)/\{(ln$^{2}$($T$/$T_{K}$)+$\pi ^{2}S$($S$+1)\}$^{1/2}$], where $\rho
_{imp}$ is an temperature-independent impurity scattering resistivity, $\rho
_{0}$ is proportional to the concentration of the local magnetic moment, $%
T_{K}$ is Kondo temperature, and $S$ is set as 1/2.\cite{Hamann} The fitted
parameters are $\rho _{imp}$=0.76(1) $m\Omega \cdot cm$, $\rho _{0}$=0.54(1)
$m\Omega \cdot cm$, $T_{K}$=24.3(4) K. Inset of Fig. 4(a) shows the region
at low temperature, where it can be seen that Hamann's equation is valid
approximately down to temperatures $T\sim T_{K}$. It should be noted that
after deducting the magnetoresistance, the normal-state resistivity at $\mu
_{0}H$=35 T still increases with decreasing temperature for both field
directions, showing saturation trend as expected for $\rho $($T$) of diluted
impurities below T$_{K}$.\cite{Monod} This behavior is an important
distinction from the metallic resistivity above $T_{C}$ in 1111 and 122
systems, even FeSe$_{x}$,\cite{Hsu FC}$^{,}$\cite{Yuan HQ}$^{,}$\cite{Jaroszynski}
because these systems do not contain excess Fe with local moment.

Negative magnetoresistance (NMR) in the normal state (Fig. 4(b)) further
suggests the effect of excess Fe. NMR observed in S-09 is rather unusual
when compared to other iron-based superconductors, such as 1111 and 122
systems where positive MR violates Kohler scaling due to multiband effects
or the depletion of density of states at the Fermi surface with temperature
change.\cite{Luo HQ} Observed NMR is most likely ascribed to suppressing
incoherent Kondo spin-flip scattering, which has been intensively studied in
dilute alloy systems.\cite{Monod2} The absolute values of MR increase with
increasing field at the constant temperature and the MR effect is weaker
with the temperature increase. Moreover, the MR effect is more pronounced
for H$\Vert $ab than for H$\Vert $c. Similar NMR have been seen in excess
iron doped TaSe$_{2}$ and ascribed to Kondo-type scattering.\cite{Whitney}

Fig. 4(b,c) show the magnetic-field dependence of Hall resistivity $\rho
_{xy}$ and $R_{H}$ determined from the slope of Hall resistivity $\rho
_{xy}(H)$ at different temperatures. The positive $R_{H}$ above $T_{C}$
indicates that the electronic transport is dominated by hole-type carriers.
We observe no abrupt change in carrier density at the temperature of
magnetic transition.\cite{Hu RW} This is consistent with ARPES and optical
spectroscopy observations, implying that there is no gap at the Fermi
surface below the magnetic transition. There is strong temperature
dependence of $R_{H}$ that increases continuously with increasing
temperature. The $R_{H}$ bending at low temperatures can be ascribed to skew
scattering.\cite{Fert} The combined influence of ordinary Hall effect R$_{0}$
and skew scattering due to Kondo scattering is expected to follow R$_{H}$%
(T)=R$_{0}$+A/(T-$\Theta $) dependence where $\Theta $ characterizes the
strength of exchange interaction between local moments. We obtained
excellent fitting results (red curve in Fig. 4(c)). The value of $\Theta $
is -57.9(1) K, which indicates that the exchange interaction between Fe is
antiferromagnetic. The similar behavior has also been observed in other
iron-based systems was explained by localization behavior induced by
disorder, multiband effects or partial gapping Fermi surface with decreasing
temperature.\cite{Liu TJ}$^{,}$\cite{Cheng P}$^{,}$\cite{Chen GF3}
Furthermore, using the obtained R$_{H}$($T$=0)=3.02$\times $10$^{-8}$ m$^{3}$%
/C, i.e. the zero-temperature carrier concentration $n$=2.07$\times $10$^{20}
$ cm$^{-3}$, and obtained residual resistivity $\rho _{0}$=1.84 $m\Omega
\cdot cm$ from the Hamann's equation, we can evaluate the mean free path of
S-09, $l$=1.35 nm using Drude model $l$=$\hbar (3\pi ^{2})^{1/3}/(e^{2}\rho
_{0}n^{2/3})$. This confirms that S-09 is a dirty-limit superconductor since
$l/\xi (0)$=0.396.

\section{Conclusion}

In summary, the anisotropy in the upper critical field of Fe$_{1.14(1)}$(Te$%
_{0.91(2)}$S$_{0.09(2)}$)$_{z}$ single crystals was studied in high and
stable magnetic fields up to 35 T. We found that the zero-temperature upper
critical field is much smaller than the predicted result of WHH theory
without the spin-paramagnetic effect. The anisotropy of the upper critical
field decreases with decreasing temperature, becoming nearly isotropic at
low temperature. The spin-paramagnetic effect is the dominant pair-breaking
mechanism for both H$\Vert $ab and H$\Vert $c crystallographic axes. There
is obvious spin-orbital scattering effect in this system. Our results show
no abrupt change in the carrier density at the temperature of magnetic
transition and considerable Kondo-type scattering effects on resistivity, MR
and Hall properties. All of these results indicate that the excess Fe in
Fe(2) site act as Kondo-type impurities and play a key role in the exotic
normal and superconducting state properties.

\section{Acknowledgements}

We thank Vladimir Dobrosavljevic, Ruslan Prozorov and Weiguo Yin for useful
discussions and T. P. Murphy for experiment support in NHMFL. This work was
carried out at the Brookhaven National Laboratory, which is operated for the
U.S. Department of Energy by Brookhaven Science Associates
DE-Ac02-98CH10886. This work was in part supported by the U.S. Department of
Energy, Office of Science, Office of Basic Energy Sciences as part of the
Energy Frontier Research Center (EFRC), Center for Emergent
Superconductivity (CES). A portion of this work was performed at the
National High Magnetic Field Laboratory, which is supported by NSF
Cooperative Agreement No. DMR-0084173, by the State of Florida, and by the
U.S. Department of Energy.

$^{\ast }$Present address: Ames Laboratory US DOE and Department of Physics
and Astronomy, Iowa State University, Ames, IA 50011, USA.

\end{document}